\documentclass[ reprint,
 amsmath,amssymb,
 aps,
]{revtex4-1}

\usepackage{graphicx}
\usepackage{dcolumn}
\usepackage{bm}

\usepackage{ulem}

\begin{document}

\title{Compact Group Approach to the Analysis of Dielectric and\\
Optical Characteristics of Finely Dispersed Systems and Liquids}
\thanks{Journal of Physical Studies \textbf{13}, No. 4 (2009) 4708
(5p)\\
Author's e-mail: mrs@onu.edu.ua}

\author{M.~Ya.~Sushko}
\affiliation{Department of Theoretical Physics, Mechnikov National
University,  2 Dvoryanska St., Odesa 65026, Ukraine}


\begin{abstract}
We present the compact group approach to analysis of the
long-wavelength dielectric and optical characteristics of
substances which can be modeled as macroscopically homogeneous and
isotropic systems of hard dielectric particles embedded into a
dielectric matrix. After the basics of the method are outlined, we
discuss the results obtained with it for the effective
permittivity of such systems and the single-scattering intensity
of light scattered by them.

\end{abstract}



\maketitle

\section{Introduction}

The dielectric and optical properties of particulate substances, such as dielectric composites, suspensions, and liquids, are often studied in terms of the simplest and widely-spread model which treats these substances as macroscopically homogeneous and isotropic systems of hard dielectric inclusions embedded into a host medium of constant permittivity. A rigorous analysis of it with methods of the multiple scattering theory requires that an infinite set of the correlation functions of the system be known. Because of the lack of this information, practical implementations of the multiple scattering approach are usually limited to the first Born approximation or heavily rely on uncontrolled assumptions about many-particle correlations and multiple scattering effects within the system. In addition, direct attempts at going beyond the limits of the first Born approximation inevitably face mounting computational difficulties, including divergences in higher-order terms of the iterative series (see, for instance, an extensive review \cite{Tsang2001} and recent works \cite{Mallet2005,Kuzmin2005}). The discrepancy between the theory and experiment can be as high as $30\%$.

In this report, we present a new method \cite{SushkoJETP:2007} for studying the long-wavelength dielectric and optical characteristics of macroscopically homogeneous and isotropic particulate systems, which is based upon the concept of macroscopic compact groups of particles. These groups  are defined as macroscopic regions that have typical scales much smaller than the wavelength of probing radiation in the host, but yet include sufficiently large numbers of particles to reproduce the properties of the entire system. In the particular case of liquids, compact groups can be interpreted as small physical volumes, containing macroscopic numbers of atoms (molecules).

In the long-wavelength limit, the fluctuations of particle numbers in compact groups are negligibly small. On the other hand, compact groups can be treated with respect to a long-wavelength field as one-point inhomogeneities. Correspondingly, a particulate system can be viewed as a set of such groups, with the local value of the permittivity tensor in the system given by $\varepsilon_{ik}(\textbf{r}, t) = \varepsilon_{0}\delta_{ik}+ \delta\varepsilon_{ik} (\textbf{r}, t)$, where $\varepsilon_{0}$ is the permittivity of the host medium and the term $\delta\varepsilon_{ik} (\textbf{r}, t)$ is due to the presence of a compact group of particles  at point $\textbf{r}$ at instant $t$. The explicit form of $\delta\varepsilon_{ik} (\textbf{r}, t)$ is modeled in accordance with the type of particles constituting the system.

The concept of compact groups allows us to avoid excessive refinement of short-range multiple scattering processes and interparticle correlations inside the system. Their averaged contributions can be effectively singled out from all terms of the iterative series for the electric field and displacement and then summed up. In this report, we  use the compact group approach to analyze the dielectric and optical characteristics of a system of hard inhomogeneous dielectric balls whose scalar permittivity is given by a  piecewise-continuous function $\varepsilon=\varepsilon(\textbf{r})$. The necessary details of the approach are presented in section 2. The effective dielectric constant of the system is calculated in section 3. The intensity of single scattering (in a generalized sense) of light from the system is discussed in section 4.

\section{Basic Relations}

Consider a macroscopically homogeneous and isotropic system of hard (nonoverlapping) dielectric balls with radius $R$ and scalar permittivity $\varepsilon(\textbf{r})$. Suggesting that the contribution from a compact group is a scalar quantity, we represent the local permittivity distribution in the system as
\begin{equation} \label{a1}
\varepsilon (\textbf{r}, t) = \varepsilon_{0} + \delta \varepsilon (\textbf{r},t),
\end{equation}
\begin{equation} \label{a2}
\delta \varepsilon (\textbf{r},t) =\sum_{a =1}^{N} \Delta\varepsilon ({\textbf{r} - \textbf{r}_{a}(t)) \theta (R - |\textbf{r} - \textbf{r}_{a}(t)|)},
\end{equation}
where  $\Delta\varepsilon(\textbf{r})= \varepsilon(\textbf{r}) -\varepsilon_{0}$, $\theta(x)$ is the Heaviside function, and the summation is carried out over the position vectors $\textbf{r}_{a} = \textbf{r}_{a}(t)$ of the centers of $N$ balls, making up a compact group.

For model \eqref{a1}--\eqref{a2} and relatively slow time variations of $\delta \varepsilon ({\textbf{r}},t)$, the equation describing the propagation of an electromagnetic wave within such a system can be represented as \cite{Landau:1982, SushkoLTP:2007}
\begin{equation} \label{a3}
\Delta \textbf{E} + k_{0}^{2} \varepsilon _{0} \textbf{E} -
\textrm{grad} \textrm{div} \textbf{E} = - k_{0}^{2} \delta
\varepsilon \textbf{E},
\end{equation}
where $\textbf{E}(\textbf{r},t)$ is the wave electric field and $k_{0} =2\pi/\lambda_{0}$ is the wavevector in vacuum. Changing to the equivalent integral equation and solving it by the iteration method, we represent the formal solutions for the local electric field and displacement in the form of infinite iterative series.

The compact-group contributions in these iterative series are formed by those ranges of coordinate values where the electromagnetic field propagators reveal a singular behavior. For the electromagnetic field propagator acting on a set of scalar, compactly supported, and bounded functions $\delta \varepsilon (\textbf{r},t)$, we derive the representation \cite{SushkoJETP:2004} (see also \cite{Beresteckii:1982} for the particular case $k_{0} \rightarrow 0$)
\begin{equation} \label{a4}
\tilde {T}_{\alpha \beta}  (r) =  {\frac{{1}}{{3k^{2}}}}\delta ({\rm {\bf
r}})\delta _{\alpha \beta}  \,e^{ikr}  +
{\frac{{1}}{{4\pi k^{2}}}}\left(
{{\frac{{1}}{{r^{3}}}} - {\frac{{ik}}{{r^{2}}}}} \right)
\end{equation}
$$\times \left( {\delta _{\alpha \beta}  - 3e_{\alpha}  e_{\beta}}
\right)\,e^{ikr} -
 {\frac{{1}}{{4\pi r}}}\left( {\delta _{\alpha \beta}
 - e_{\alpha}  e_{\beta}}   \right)\,e^{ikr},
$$
where $k=k_{0}\sqrt{\varepsilon_{0}}$ is the wavevector in the host, $\delta_{ik}$ is the Kronecker delta, $\delta(\textbf{r})$ is the Dirac delta function, and $e_{\alpha}$ is the $\alpha$ component of the unit vector $\textbf{e}=\textbf{r}/r$. Using \eqref{a4}, the compact-group contributions can be singled out from all terms in the iterative series by replacing all inner propagators by their most singular parts (the first term in \eqref{a4}). We suggest that effects of multiple reemissions and short-range correlations within compact groups predominate in the formation of the long-wavelength dielectric and optical characteristics of the system. For problems dealing with the quasistatic effective permittivity, this statement can be proven rigorously \cite{SushkoJETP:2007, SushkoArxiv:2009}.

\section{Quasistatic Effective permittivity}

We define the quasistatic effective permittivity $\varepsilon_{\rm{eff}}$ of a particulate system by the relation
\begin{equation} \label{b1}
\langle \textbf{D} (\textbf{r})\rangle = \langle \epsilon (\textbf{r}) \textbf{E} (\textbf{r}) \rangle = \varepsilon_{\rm{eff}} \langle \textbf{E} (\textbf{r}) \rangle,
\end{equation}
where $\textbf{D} (\textbf{r})$, $\textbf{E} (\textbf{r})$, and $\epsilon (\textbf{r})$  are the local values of,  respectively, the electric displacement, electric field, and permittivity in the system. The angle brackets in \eqref{b1} stand for the averages, which, depending on the problem to be solved, can be calculated using either the formalism of many-particle correlation functions \cite{Bogolubov:1946} or the averaging \cite{Landau:1982} over volumes much greater than the scales of compact groups. For finely dispersed particulate systems, both approaches are expected to give equal results.

Using the compact group approach, we prove \cite{SushkoJETP:2007, SushkoArxiv:2009} that in the long-wavelength limit $k_{0}\sqrt{\varepsilon_{0}}\rightarrow 0$, the averaged (in either way) electric field and displacement  are
\begin{equation} \label{b2}
\langle{\rm {\bf  {E}}}\rangle = {\left[ {1 + {\sum\limits_{s = 1}^{\infty}  {\left( {
- {\frac{{1}}{{3\varepsilon _{0}}} }} \right)^{s} \langle {{\mathop {\left( {\delta
\varepsilon ({\rm {\bf r}})} \right)^{s}}}}}
}}\rangle \right]}\,{\rm {\bf E}}_{0},
\end{equation}
\begin{equation}\label{b3}
\langle{\rm {\bf {D}}}\rangle = {\varepsilon _{0}\left[ {1 - 2
{\sum\limits_{s = 1}^{\infty} {\left( { -
{\frac{{1}}{{3\varepsilon _{0} }}}} \right)^{s}{\langle {\mathop
{\left( {\delta \varepsilon ({\rm {\bf r}})}
\right)^{s}}}}\rangle} } }  \right]}\,{\rm {\bf E}}_{0},
\end{equation}
where $\textbf{E}_{0}$ is the amplitude of the incident wave field in the host.

The statistical averages in \eqref{b2} and \eqref{b3} are found readily if we take into account that for a macroscopically homogeneous and isotropic system,  its one-particle correlation function $F_{1}(\textbf{r}_1)=1$ and its many-particle correlation functions  $F_{s}(\textbf{r}_{1},...,\textbf{r}_{s})$, $s \geq 2$, depend only on the mutual interparticle distances $|\textbf{r}_{i}-\textbf{r}_{j}|$ ($i, j =1,2,..., s$, $i\neq j$). Moreover, for a system of hard (nonoverlapping, but in general interacting) balls, every $F_{s}$, $s \geq 2$, vanishes for any particle configuration in which at least one of the distances  $|\textbf{r}_{i}-\textbf{r}_{j}|$ is shorter than $2 R$. Together with \eqref{a2}, these properties of $F_{s}$ immediately give
\begin{equation}\label{b4}
{\langle {\mathop {\left( {\delta \varepsilon ({\rm {\bf r}})}
\right)^{s}}}\rangle}  =  n {\int\limits_{\Omega} {d{\rm {\bf
r}}\, {\left( \Delta \varepsilon ({\rm {\bf{r}}}) \right)}^{s}}},
\quad s \ge 1,
\end{equation}
where $n =N/V$ is the particle concentration and the integral is taken over the region $\Omega$ occupied by one particle. In view of \eqref{b4}, expressions \eqref{b2} and \eqref{b3} are easy to sum up. With definition \eqref{b1}, the effective permittivity of the system can be represented in the form
\begin{equation}\label{b5}
\varepsilon_{\rm{eff}} =\varepsilon_{0} \frac {1 + {\frac {8 \pi }{3}} n \alpha }
{1 - {\frac {4\pi}{3}}n \alpha},
\end{equation}
where
\begin{equation} \label{b6}
\alpha  = \frac{3}{4 \pi} {\int \limits_{\Omega}} d {\rm {\bf r}}\,
\frac {\varepsilon (\textbf{r}) - \varepsilon_{0}} {\varepsilon(\textbf{r}) + 2 \varepsilon_{0}}.
\end{equation}

Expression \eqref{b5} coincides with the well-known Lorentz--Lorenz formula \cite{Born:1969,Frohlich:1958}, provided the quantity $\alpha$ is interpreted as the one-particle polarizability of balls in the system. Obviously, this result counts in favor of the compact group approach. It should also be emphasized that, in contrast to the standard case of homogeneous particles, we derived it for a system of inhomogeneous (say, layered) dielectric balls. In doing so, no restriction on the value of the difference $\Delta \varepsilon(\textbf{r})$ was imposed. In the case of homogeneous balls, formulas \eqref{b5} and \eqref{b6} immediately result in the well-known Maxwell-Garnett mixing rule (see, for instance, \cite{Sihvola:1999}). That property of particles that they are hard is evidently behind the appearance of this rule. This our conclusion is different from that in \cite{Mallet2005}.

The same results are obtained \cite{SushkoJETP:2007} if the averages in \eqref{b1}--\eqref{b3} are calculated using the  averaging procedure \cite{Landau:1982}, that is, integration over volumes much greater than the scales of compact groups. In this way, generalizations of formulas \eqref{b5}, \eqref{b6}, and the Maxwell-Garnett mixing rule are obtained  in \cite{SushkoJTP:2009,SushkoArxiv:2009} for multicomponent systems of hard anisotropic inhomogeneous dielectric particles.

\section{Single-Scattering Light Intensity}
\subsection {Light Scattering from Liquids}
The traditional approach  \cite{Komarov:1962} to finding the intensity of light scattered by fluids far enough from the critical point is limited to the first Born approximation in the multiple scattering theory. This implies that the scattering intensity is formed by the effects involving only pairs of scattering particles. Correspondingly, there is no fundamental difference between the ways in which dense liquids and rarefied gases are treated, despite the fact that mutual polarization effects of neighboring particles on one another are obviously more significant in liquids. Let us show that the compact group approach makes it possible to fix this defect of the theory.

Physically, the scattering by a pair of compact groups of particles (that is, by a pair of small physical volumes) should be treated as single. The general expression for the overall single-scattering spectrum due to scattering from all such pairs can be represented as (see \cite{SushkoLTP:2007})
\begin{equation} \label{d2}
I(\textbf{q}, \omega)= \sum_{r,\, s = 1}^{\infty} {I_{rs}(\textbf{q}, \omega)},
\end{equation}
where
\begin{equation} \label{d3}
I_{rs} (\textbf{q}, \omega) \propto \left(- \frac{1}{3 \varepsilon_{0}}\right)^{r+s-2} \frac{1}{\pi}
\end{equation}
$$ \times  \,
\textrm{Re} \int \limits_{0}^{\infty} dt \int \limits_{V}
d\textbf{r} \langle (\delta {\varepsilon(\textbf{r},t)})^{r}
(\delta \varepsilon(\textbf{0},0))^{s} \rangle e^{i\omega t - i
\textbf{q}\cdot\textbf{r}}
$$
is the contribution from $r$ and $s$ acts of scattering inside a pair of compact groups; $\textbf{q}$ and $\omega$ are the changes in the light wavevector and frequency due to scattering, respectively. The proportionality factor in \eqref{d3} is $I_{0} V k_{0}^{4} \textrm{sin}^{2} \gamma/16 \pi^2 R_{0}^2$, where $I_{0}$ is the incident wave intensity, $V$ is the scattering volume, and $\gamma$ is the angle between the polarization vector of the incident wave and the direction at the observation point at a distance $R_{0}$ away.

Using formulas \eqref{a1}, \eqref{a2}, \eqref{d2}, and \eqref{d3}, and taking into account the macroscopic homogeneity and isotropy of the system, the summation in \eqref{d2} can be carried out to give
\begin{equation} \label{d4}
I(\textbf{q}, \omega) =   I_{0} V \frac{  k_{0}^{4} \textrm{sin}^{2} \gamma  }{2 \pi R_{0}^2} \varepsilon_{0}^{2} |\alpha(\textbf{q})|^{2} n G(\textbf{q}, \omega),
\end{equation}
where  $G(\textbf{q},\omega)$ is the frequency Fourier transform
of the  Van-Hove function
\begin{equation} \label{d5}
G(\textbf{q},t) = \frac{1}{N} \langle {\sum_{a =1}^{N} e^{-i\textbf{q}\cdot\textbf{r}_{a}(t)}}
{\sum_{b =1}^{N} e^{i\textbf{q}\cdot\textbf{r}_{b}(0)}} \rangle
\end{equation}
for the particles within a compact group,
and
\begin{equation} \label{d6}
\alpha(\textbf{q})= {\frac{3}{4 \pi}} \int \limits_{\Omega} d \textbf {r} \frac{\varepsilon(\textbf{r})-\varepsilon_{0}}{\varepsilon (\textbf{r})+ 2 \varepsilon_{0}} e^{-i \textbf{q} \cdot\textbf{r}}.
\end{equation}
The integrated intensity of light scattered at nonzero scattering angles $\theta$ ($\textbf{q} \neq 0)$ is
\begin{equation} \label{d7}
I(\textbf{q}) =    I_{0} V \frac{k_{0}^{4} \textrm{sin}^{2} \gamma  }{R_{0}^2}\varepsilon_{0}^{2}  |\alpha(\textbf{q})|^{2}n S(\textbf{q}),
\end{equation}
where $S(\textbf{q})$ is the static structure factor of the liquid.

For macroscopic compact groups of particles interacting by means of short-range potentials, the function \eqref{d5} is evidently equal to the Van-Hove function for the entire liquid. Also, in the limit $\textbf{q}\rightarrow 0$, the quantity \eqref{d6} becomes equal to the effective polarizability \eqref{b6} of a particle in the liquid (at the frequency of the incident wave). It follows that multiple short-range reemissions between the particles within compact groups result in the renormalization of the effective polarizability of the particles in the liquid, but do not affect the spectral and angular distributions of the scattered light intensity as compared to the well-known result \cite{Komarov:1962}, obtained within the first Born approximation.

The applications of formulas  \eqref{d2} and \eqref{d3}, with different interpretation of the concept of compact    groups, to the problem of so-called 1.5-scattering of light by fluids near the critical point can be found in \cite{SushkoJETP:2004, SushkoLTP:2007}.

\subsection {Light Scattering from Suspensions}
It is evident that reasoning of subsection 4.1 also applies to finely dispersed suspensions. To put our result to the test, we contrast it with the analytical result \cite{Kuzmin:2001} for a suspension of homogeneous hard balls. The authors of \cite{Kuzmin:2001} used a different way to find the integrated intensity $I({q})$. Namely, exploiting the exact solution \cite{Wertheim:1963,Thiele:1963} of the Percus--Yevick integral equation for the radial distribution function for hard balls, they calculated rigorously the pair correlation function for  fluctuations of the static local permittivity given by \eqref{a1} and \eqref{a2} ($t=0$); then they found $I({q})$ in the first Born approximation.

For homogeneous hard balls of radius $R$ and the Percus--Yevick radial distribution function \cite{Wertheim:1963,Thiele:1963}, we have
\begin{equation} \label{d8}
\alpha(q)=3 \frac {\varepsilon - \varepsilon_{0}}{\varepsilon+2\varepsilon_0} \frac{R}{q^{2}} \left( \frac{\textrm{sin} qR}{qR} - \textrm{cos} qR \right),
\end{equation}
\begin{equation} \label{d9}
S(q)= \frac{1}{1-n c(q)},
\end{equation}
where $c(q)$ is the Fourier transform of the direct correlation function. Substituting these formulas into \eqref{d7}, we find that our expression for $I(q)$ is almost the same as that in \cite{Kuzmin:2001}. The only difference is the factor $ K \equiv 9 \varepsilon_{0}^2/(\varepsilon + 2\varepsilon_{0 })^{2}$, which appears in our formula. In particular, $K<1$ if $\varepsilon > \varepsilon_{0}$. We conclude that multiple short-range reemissions between particles in such a concentrated suspension effectively reduce the inhomogeneity effects induced by single particles; as a result, the scattering mean free path $l$ and the transport mean free path $l^{*}$ increase.

As a further check, we calculated $l$ and $l^{*}$ by the formulas
\begin{equation} \label{d10}
\frac{1}{l}= \frac{k_{0}^{4}\varepsilon_{0}^{2}}{2}\int d\Omega \left( 1+ \textrm{cos}^{2}\theta\right)|\alpha(\textbf{q})|^{2} n S(\textbf{q}) ,\,\, \,\,\,\, l^{*}=\frac{l}{1-g},
\end{equation}
where the scattering anisotropy parameter $g$ is the average of $\textrm{cos} \theta$,
\begin{equation} \label{d11}
g = \overline {\textrm{cos} \theta } = \frac {\int d\Omega \left( 1+ \textrm{cos}^{2}\theta\right)\textrm{cos}\theta |\alpha(\textbf{q})|^{2} n S(\textbf{q})}{\int d\Omega \left( 1+ \textrm{cos}^{2}\theta\right) |\alpha(\textbf{q})|^{2} n S(\textbf{q})},
\end{equation}
the integrals are taken over the entire solid angle, $S(\textbf{q})$ is given by \eqref{d9} with
($d=2R$)
\begin{eqnarray} \label{d12}
c(q)=\frac{4 \pi d}{q^{2}} \left[ (\alpha+\beta+\delta) \left(\textrm{cos} q d - \frac{\textrm{sin} q d}{q d} \right)
 \right.\quad\quad  \nonumber \\
\left.-\beta \left(\frac {\textrm{sin} q d}{q d}+ 2 \frac{\textrm{cos} q d -1}{q^{2} d^{2}} \right) \right.\quad\quad \quad\quad \nonumber \\
\left.- \delta \left(3\frac {\textrm{sin} q d}{q d}+ 12
\frac{\textrm{cos} q d}{q^{2}d^{2}} \right. \left. - 24 \frac
{\textrm{sin} q d}{q^{3}d^{3}}-24 \frac{\textrm{cos} q d-1}{q^{4}
d^{4}} \right) \right], \nonumber\\
\end{eqnarray}
\begin{equation} \label{d13}
\alpha =\frac{(2\eta+1)^{2}}{(1-\eta)^{4}},\,\, \beta = -\frac{6\eta(\eta/2+1)^{2}}{(1-\eta)^{4}},\,\,\, \delta = \frac {\alpha \eta}{2},
\end{equation}
and $\eta=4\pi R^{3} n/3$ is the volume concentration of dispersed
particles.

\bigskip
\begin{table}[h]
\begin{tabular}{ccccccc}
\hline
  $\eta$      & $\overline{N}$   & $l, \mu \rm{m}$   &    &
  $\overline{N}$  & $l^{*}, \mu \rm{m}$   & \\

              &                  & Exp.              & Eq.~\eqref{d10}
  &                & Exp.             & Eq.~\eqref{d10} \\
\hline
0.026 &      &       &       & 6.1   & 570  & 497\\
0.037 & 6.2  & 227   & 227   & 8.8   & 396  & 367\\
0.053 &      &       &       & 12.6  & 283  & 281\\
0.054 & 9.0  & 180   & 172   &       &      &     \\
0.076 & 12.8 & 140   & 139   & 18.0  & 225  & 224\\
0.11  & 18.6 & 125   & 121   & 26.2  & 214  & 191\\

\hline
\end{tabular}
\caption{Scattering mean free path $l$ and transport mean free
path $l^{*}$ for aqueous suspensions of latex particles according
to experimental data \cite{Saulnier:1990} and present theory. }
\label{tab:a}
\end{table}
\noindent

The results are presented in Table \ref{tab:a} and compared with
experimental data \cite{Saulnier:1990} for aqueous suspensions of
latex particles ($\sqrt{\varepsilon_{0}} =1.33$,
$\sqrt{\varepsilon} =1.59$, $d=0.087 \mu{\rm m}$); the paths $l$
and $l^{*}$ were measured in \cite{Saulnier:1990} using light with
$\lambda_{0}$ = $0.5145 \mu {\rm m}$ and $0.5785 \mu {\rm m}$,
respectively. The average numbers of particles per cubic
wavelength, $\overline{N} = n \lambda^{3}$, where $\lambda$ is the
wavelength in water, are also shown. As can be seen, the data
\cite{Saulnier:1990} and our results are in a surprisingly good
agreement.

\end{document}